\documentclass[tightenlines,superscriptaddress,eqsecnum,floats,nofootinbib,showpacs]
{revtex4}
\usepackage{amssymb}
\usepackage{stmaryrd}
\usepackage{amsmath}
\usepackage{amsfonts}
\usepackage{mathrsfs}
\usepackage{amsmath,amssymb,amsfonts}
\usepackage{graphicx}
\usepackage[T1]{fontenc}
\usepackage{color} 

\def\be{\begin{equation}}
\def\ee{\end{equation}}
\def\ba{\begin{eqnarray}}
\def\ea{\end{eqnarray}}
\def\nn{\nonumber}

\newcommand{\sgn}{\mathrm{sgn}} 
\newcommand{\dd}{{\rm d}}

\begin{document}


\title{ Loop quantum black hole}

\author{Xiangdong Zhang\footnote{scxdzhang@scut.edu.cn}}
\affiliation{Department of Physics, South China University of
Technology, Guangzhou 510641, China}

\begin{abstract}
In the last decades, progress on the quantization of black holes using techniques developed in loop quantum cosmology has received increasing attention. Due to the quantum geometry effect, the resulting quantum corrected black hole is free of singularity. The quantization scheme can be roughly divided into four types, that is 1. $\mu_0$-scheme, 2. $\bar{\mu}$-scheme, 3. generalized $\mu_0$-scheme, 4. quantum collapsing model. This paper provides an introduction of the loop quantum black hole models and a summary of the progress made in this field, as well as the quantum effective dynamics and physical applications of these models.

\end{abstract}


\maketitle

\section{Introduction}
Building a consistent model to unify quantum mechanics and general relativity (GR) is a holy grail to the current theoretical physicist. Loop quantum gravity (LQG) serves as one of the promising candidates for quantum gravity which is featured by its nonperturbative characteristics \cite{Rovelli:2004tv,Thiemann:2007pyv,Ashtekar:2004eh,Han:2005km}. Over the past years, LQG achieves a series of notable progress including making the natural predictions of the discrete geometrical spectrum and providing a reasonable microscopic interpretation of Hawking-Beikenstein black hole entropy \cite{Rovelli:1994ge,Ashtekar:1996eg,Ashtekar:1997fb,Yang:2016kia,Thiemann:1996at,Ma:2010fy,Ashtekar:1997yu,Song:2020arr,Perez17}. We {refer} \cite{Perez17} for complete review of the results on black hole physics in the framework of loop quantum gravity. This nonperturbative LQG quantization framework recently has also been extended to the metric $f(R)$ theories, Weyl gravity, scalar-tensor theories as well as higher-dimensional gravity and so on \cite{Zhang:2011vi,Zhang:2011qq,Zhang:2011gn,Zhang:2011vg,Bodendorfer:2011nv,Zhang:2020smo}. Despite these remarkable achievements, the LQG dynamics are still not fully understood. In order to understand the dynamics of LQG, one often utilizes symmetry-reduced models to test the quantization technologies developed in full LQG. Particularly to the Friedmann-Robertson-Walker (FRW) universe model, leading to the field of loop quantum cosmology (LQC) \cite{Ashtekar:2003hd,Ashtekar:2005qt,k=-1LQC}. The most striking feature of LQC is it can naturally replace the classical big bang singularity of the universe with a quantum bounce and resulting a nonsingular evolution of the universe. We refer to \cite{Ashtekar:2003hd,Bojowald:2005epg,Ashtekar:2006wn,Ashtekar:2011ni} for more complete reviews on LQC.

In addition to the big-bang singularity in cosmological models, another well-known singularity lies at the center of a black hole. For instance, the simplest model is the Schwarzschild interior space-time which contains a singularity at $r=0$. The interior of a Schwarzschild black hole can be isometric to the Kantowski-Sachs model \cite{AB06,Vandersloot07}. Thus, the techniques developed for LQC can naturally transport to the spherically symmetric Schwarzschild black hole model. This leads to the field of so-called loop quantum black hole models, we refer to \cite{AB06,Vandersloot07,Chiou08} for detailed constructions of the model. Like LQC can solve the big bang singularity of the universe. In the loop quantum black hole model, the black-hole interior singularity can also be resolved as expected. We would like to emphasize that in review \cite{Perez17} concerns of the results on black hole physics in the framework of loop quantum gravity, while in {this paper}, we focus on the loop quantum black hole models utilizing the techniques developed in loop quantum cosmology.

Moreover, in contrast to the LQC which has a consistent treatment for various models. In the loop quantum black hole model, different models usually choose different quantum parameters to regularize and quantize the Hamiltonian constraint. Generally, these loop quantum black hole models can be divided into the following four schemes. The first strategy is the $\mu_o$-scheme. In this scheme, the quantum regularization parameters are set to be constants by certain considerations \cite{AB06}. The second strategy is the $\bar\mu$-scheme \cite{Vandersloot07,Chiou08} with the quantum regularization parameters chosen as a function of the phase space variables. The third one is the so-called generalized $\mu_0$-scheme \cite{AOS1}. And the fourth type is the quantum collapsing model \cite{Yang22}. Although the singularity resolution of the Schwarzschild interior holds in both of these methods, the detailed construction of the effective dynamics is quite different. In some schemes, inconsistent physical results will occur. For instance, in $\mu_o$-scheme scheme, the quantum bounce which replaces the classical singularity of Schwarzschild interior could appear in the low curvature region \cite{AB06,M06,CGP08}. While in $\bar\mu$-scheme the quantum corrections to the Schwarzschild black hole horizon could be large, however, since the curvature in horizon usually very small, one generally believed that horizon can be considered as the classical region and should not be received too large quantum corrections \cite{Vandersloot07,Chiou08}. In attempts to cure these weaknesses, recently some authors proposed new schemes generalized $\mu_0$-scheme as well as the quantum collapsing model \cite{AOS1,AOS2,Yang22}. Moreover, as suggested by some authors \cite{Ewing20}, it is maybe not suitable to use
the $\bar{\mu}$-scheme near the horizon since a spatial coordinate will become null at
the horizon. Hence they suggest implementing the $\bar{\mu}$-scheme in a set of {special}
(spatial) coordinates {which} will not become
null at {the} horizon. By utilizing the areal gauge and using the Painleve-Gullstrand coordinates, they obtain a quantum corrected Schwarzschild metric which has the correct semi-classical limit \cite{Ewing20}.

{Due to all these interesting developments in the past decades.} The aim of this paper is thus to give a summary of these progresses as well as the quantum effective dynamics and physical applications of these {loop quantum black hole} models.

We organize the paper as follows: We first recall the Hamiltonian framework of the classical Schwarzschild black hole interior in section \ref{sec:two}. Then we study the quantum effective dynamics in section \ref{sec:three}, with chosen the different types of quantum parameters. The physical applications of the loop quantum black hole are discussed in section \ref{section4}. The conclusions and outlooks are summarized in section \ref{sec:con}. Throughout the paper we {adopt} the convention $c=1$.

\section{Classical theory}\label{sec:two}

\subsection{Preliminaries}

The action of GR reads
\begin{equation}
{S=\frac{1}{16\pi G}\int \dd^4 x\sqrt{-g}\mathcal{R},}
\end{equation} {where $g$ is the determinant of spacetime metric $g_{ab}$ and $\mathcal{R}$ denotes its scalar curvature.}
Begin with this action, and choose the spatial metric $q_{ab}$ and its conjugate momentum $p^{ab}$ as canonical variables. We can cast the whole system into the geometrical dynamics with the Hamiltonian as follows \cite{Thiemann:2007pyv}:
\ba
H_{grav}=\int d^3x \left(\mathbb{C}_aN^a+CN\right),
\ea where the diffeomorphism and Hamiltonian constraints read \ba
\mathbb{C}_a&=&D^bp_{ab}=0,\\
C&=&\frac{2\kappa}{\sqrt{q}}\left(p_{ab}p^{ab}-\frac12p^2\right)-\frac{\sqrt{q}}{2\kappa}R=0,
\ea here $q$ is the determinant of metric $q_{ab}$, {and R represents scalar curvature of $q_{ab}$} and $\kappa=8\pi G$. We can further cast this Arnowitt-Deser-Misner (ADM) Hamiltonian formalism into a connection {dynamics} formalism. By performing the canonical transformation, we introduce $E_i^a(x)=\sqrt{q}e^a_i$ with $e^a_ie^{bi}=q^{ab}$ denotes the densitied triad and $A_a^j(x)$ is its conjugated connection as the canonical variables. Since we use triad rather than metric, the additional Gaussian constraint $\mathbb{G}_i$ emerged. Moreover, the diffeomorphism and Hamiltonian constraints should also be re-expressed in terms of new variables $(A_a^j(x),E_i^a(x))$. Hence the whole Hamiltonian reads \cite{Thiemann:2007pyv}\ba
H_{grav}=\int d^3x \left(\mathbb{G}_i\Lambda^i+\mathbb{C}_aN^a+CN\right),
\ea where
\begin{equation}\label{eq:Hamilc}
\begin{aligned}
\mathbb{G}_i=&\partial_bE^b_i+\epsilon_{ij}^{\ \ k}A_a^jE^a_k=0,\\
\mathbb{C}_a=&\frac{1}{\kappa\gamma}E_j^bF^j_{ab}=0,\\
C=&\frac{1}{2\kappa}\frac{E^a_iE^b_j}{\sqrt{q}}\epsilon^{ij}_{\ \ k}\left(F^k_{ab}-(1+\gamma^2)\epsilon^k_{\ mn}K_a^mK^n_b\right)=0,
\end{aligned}
\end{equation}
represent respectively the Gaussian, spatial diffeomorphism and Hamiltonian constraints. Moreover, $\gamma$ denotes the Immirzi parameter \cite{Thiemann:2007pyv}. In addition, $F^i_{ab}=\partial_aA^i_b-\partial_bA^i_a+\varepsilon^i_{jk}A^j_aA^k_b$ and $K_a^i$ relates to the extrinsic curvature $K_{ab}$ through $K_a^i=K_{ab}e^{bi}$.

Classically, the spherically static solution to GR was the so-called Schwarzschild solution and the spacetime metric reads
\ba\label{eq:standardJNW}
ds^2=-\left(1-\frac{2GM}{r}\right) dt^2+\left(1-\frac{2GM}{r}\right)^{-1}dr^2+r^2d\Omega^2,
\ea
where $M$ stands for the Arnowitt-Deser-Misner (ADM) mass of the compact object. When {$2GM<r$}, this solution represents the exterior region of the Schwarzschild  spacetime. Contrary, when {$2GM>r$,} the Schwarzschild interior metric reads
\begin{equation}\label{eq:modifiedJNW}
\dd s^2=- \left(\frac{2GM}{r}-1\right)^{-1}\dd r^2+ \left(\frac{ 2GM}{r}-1\right)\dd x^2+r^2\dd\Omega^2.
\end{equation}
Since in this case, the killing vector $(\frac{\partial}{\partial r})^a$ becomes timelike and hence we can use the radius $r$ to represent the evolution. {The Schwarzschild spacetime has} two well-known singularities, one is the true singularity at $r=0$ and the other one is located at $r=2GM$ which can be removed by suitable coordinate transformations.

{It is easy to see that in the Schwarzschild spacetime,} the homogeneous spatial Cauchy slices $\Sigma$ possess topology $\mathbb{R}\times \mathbb{S}^2$. To manifest this symmetry, one usually introduces the following fiducial metric $\mathring{q}_{ab}$ on Cauchy slices $\Sigma$
\begin{equation}
\mathring{q}_{ab}\dd x^a\dd x^b=\dd x^2+r_o^2d\Omega^2.
\end{equation}
Here $x\in(-\infty,\infty)$, and constant $r_o$ has dimensions of length. Note that in the direction $x$, the spatial slice $\Sigma$ is non-compact. Hence an elementary cell $\mathcal{C}\cong (0,L_0)\times \mathbb{S}^2$ with a finite $L_0$ should be introduced in slice $\Sigma$ and we then calculate all integrals with respect to this elementary cell rather than the divergent {$x$-direction} to avoid the possible divergence problems.

The connection $A_a^i(x)$ and triad $E^a_i(x)$ will be greatly simplified due to the symmetry as follows \cite{Ashtekar:2005qt}
\ba
E_i^a\tau^i\partial_a&=p_c\tau_3\sin\theta\frac{\partial}{\partial x}+\frac{p_b}{L_0}\tau_2\sin\theta\frac{\partial}{\partial \theta}-\frac{p_b}{L_0}\tau_1\frac{\partial}{\partial \phi},\label{connection}\\
A_a^i\tau_i\dd x^a&=\frac{c}{L_0}\tau_3\dd x+b\tau_2\dd\theta-b\tau_1\sin\theta\dd\phi+\tau_3\cos\theta\dd\phi,\label{triad}
\ea where the $\tau_i$ are a basis of
the $su(2)$ Lie algebra.
The non-vanishing Poisson brackets between canonical variables now read
\begin{equation}
\{p_b,b\}=-G\gamma, \quad \{c,p_c\}=2G\gamma.
\end{equation}
With the help of Eqs. \eqref{connection} and \eqref{triad}, we can {also} express the curvature $F^i_{ab}$ and the extrinsic curvature $K^i_a$ as follows:
\ba
F_{ab}^i\tau_i\dd x^a\dd x^b&=&\frac{bc}{L_0}\tau_1\dd\theta\wedge \dd x+\frac{bc\sin\theta}{L_0}\tau_2\dd\phi\wedge\dd x+(-b^2\sin\theta+\sin\theta)\tau_3\dd\phi\wedge\dd\theta,\label{Fterm}\\
\gamma K_a^i\tau_i\dd x^a&=&\frac{c}{L_0}\tau_3\dd x+b\tau_2\dd\theta-b\tau_1\sin\theta\dd\phi.\label{Kterm}
\ea
The Gaussian $\mathbb{G}_i$ constraint and spatial diffeomorphism constraint $\mathbb{C}_a$ vanish automatically. We only need to consider the Hamiltonian constraint. By using Eqs.(\ref{Fterm}) and (\ref{Kterm}), the smeared Hamiltonian constraint \eqref{eq:Hamilc} now reduces to \cite{Loop_corichi_2016,Zhang2020}
\begin{equation}
\begin{aligned}
H:=\int_{\mathcal C} N C=-\frac{1}{2G\gamma}  \left(\left(b+\frac{\gamma ^2}{b}\right) p_b+2  c p_c\right),\label{HamiltonSch}
\end{aligned}
\end{equation}
here the lapse function $N$ is chosen as
\ba
N=\gamma\sgn(p_b)\frac{\sqrt{|p_c|}}{b}.
\ea
The spacetime metric with spherical symmetry then being \cite{Ashtekar:2005qt}
\begin{equation*}
\dd s^2=-N^2\dd T^2+\frac{p_b^2}{|p_c|L_0^2}\dd x^2+|p_c|\dd\Omega^2.
\end{equation*}

\subsection{The classical dynamics}
By using the Hamilton constraint \eqref{HamiltonSch}. The equations of motion read
\ba
\dot{p_c}&=&2p_c, \nn\\
\dot{c}&=&-2c,\nn\\
\dot{p_b}&=&\frac{p_b}{2}\left(1-\frac{\gamma^2}{b^2}\right),\nn\\
\dot{b}&=&-\frac12\left(b+\frac{\gamma^2}{b}\right).
\ea The solutions to the above equations {in terms of time $T$ corresponding to the lapse $N=\gamma\sgn(p_b)\frac{\sqrt{|p_c|}}{b}$ read} \cite{Loop_corichi_2016,Zhang2020}
\ba\label{eq:classicalsolution}
p_c(T)&=&4m^2 e^{2T},\\
c(T)&=&\frac{\gamma L_0}{4m}e^{-2T},\\
p_b(T)&=&-\left(e^{-T}-1\right)^{\frac12}2mL_0e^{T},\\
b(T)&=&\gamma\left(e^{-T}-1\right)^{\frac12},
\ea
where $-\infty<T\leqq0$ {and $m=2GM$.}

It is easy to see that $p_c c$ is a constant of the phase space, therefore we denotes $p_c c=: mL_0\gamma$. By comparing this solution with the schwarzschild metric \eqref{eq:standardJNW}, we can identify that $r=2me^{T}$. Moreover, singularity located at $r=0$ (or $T=-\infty$) and the black hole horizon is located at $r=2m$ (or $T=0$).

\section{Quantum theory}\label{sec:three}

To construct a viable quantum framework for Schwarzschild black holes. We adopt the standard LQC
treatment for Schwarzschild black holes which {require} to introduce of holonomy corrections. Hence in this section, we will
give a summary of the main steps and constructions.

The holonomy of an $SU(2)$-connection $A^i_a$
is path-ordering exponential integral along an edge $e^a$ \cite{Thiemann:2007pyv,Vandersloot07}
\ba
h(A):={\cal P}\exp\int_{e_i} {\rm d}t\,A^j_a\tau_j\!e^a,
\ea
with ${\cal P}$ labels path-ordering.

The effective dynamics of LQC for sharply peaked coherent states are known to provide an
excellent approximation to the full quantum dynamics \cite{Ashtekar:2006wn}. Though it is currently still not fully clear whether the effective dynamics of the loop quantum Schwarzschild black hole model will also provide a good approximation
to the full LQG dynamics for Schwarzschild black hole space-times, the success of LQC makes people believe that at least for observables whose relevant
physical length scale is much larger than Planck scale, the effective dynamics of loop quantum Schwarzschild black hole model
could also be used as a good approximation to the quantum dynamics of
semiclassical states. Based on this
expectation, we will mainly focus on the effective theory of the loop quantum black hole model here.

Inspired by LQC, a lot of models of quantum black holes have been proposed to solve the singularity inside the black hole interior. Generally \cite{Quantum_ashtekar_2005,Quantum_gambini_2014}, in the quantum effective Hamiltonian constraint, the holonomy correction is simplified by replacing the components of Ashtekar connection $b$ and $c$ with
	\begin{eqnarray}
	c\rightsquigarrow\frac{\sin(\delta_c c)}{\delta_c},\quad	b\rightsquigarrow\frac{\sin(\delta_b b)}{\delta_b},
	\end{eqnarray}
	where the quantum corrections are controlled by the quantum parameters $\delta_c$ and $\delta_b$ due to the fundamental discreteness of LQG. Thus the effective Hamiltonian of loop quantum Schwarzschild black hole can be obtained as
\ba
H_{\rm eff}=-\frac{1}{2G\gamma}\left[\left(\frac{\sin(\delta_b b)}{\delta_b}+\frac{\gamma^2\delta_b}{\sin(\delta_b b)}\right)p_b+2\frac{\sin(\delta_c c)}{\delta_c}p_c\right].
\ea

Under different choices of $\delta_b$ and $\delta_c$,  the current existing quantization schemes of the loop quantum Schwarzschild black hole can be divided into four main classes \cite{Mass_bodendorfer_2021,Properties_gan_2020}: 1. $\mu_0$-scheme\cite{Loop_modesto_2006,Semiclassical_modesto_2010}, 2. $\bar{\mu}$-scheme \cite{Loop_bohmer_2007,Quantum_alesci_2019}, 3. generalized $\mu_0$-scheme\cite{AOS1,AOS2,black_olmedo_2017,Loop_corichi_2016}, 4. quantum Oppenheimer-Snyder collapsing model \cite{Yang22}. In the following, we will discuss these four different models in more detail.

\subsection{$\mu_0$-scheme}

In the $\mu_0$-scheme, the quantum regularization parameters of $\delta_b$ and $\delta_c$ are simply taken as constants of the whole phase space \cite{AB06}. For instance, in \cite{BV07}, the author {choose} $\delta_b=\delta_c=\delta$. In loop quantum cosmology, the $\mu_0$-scheme suffers a severe problem, it will lead to a bounce in any value of matter density, therefore is unphysical. Moreover, in \cite{Singh21}, the authors show that when we consider the black hole formation in the $\mu_0$-scheme, it will suffer another drawback, that is in the $\mu_0$-scheme, unless for the unacceptable value of Barbero-Immirzi parameter, there are no trapped surfaces could be formed for a non-singular collapse of a homogeneous dust cloud in the marginally bound case.

In this scheme, a particular LQG corrected Schwarzschild
spacetime should be mentioned. The authors construct a spherically symmetric
spacetime through holonomy correction, known as the self-dual solution of LQG \cite{Loop_modesto_2006,Semiclassical_modesto_2010}. Particularly, this solution was shown that is regular and free of any spacetime curvature singularity. The metric of the self-dual spacetime by usual Schwarzschild coordinates is given by \cite{Semiclassical_modesto_2010}
\ba
\dd s^2=- g(r)\dd t^2+ f(r)\dd r^2+\left(r^2+\frac{a_0^2}{r^2}\right)\dd\Omega^2,\label{SelfdualSch}
\ea where \ba
g(r)&=&\frac{(r-r_+)(r-r_-)(r-r_*)}{r^4+a_0^2},\\
f(r)&=&\frac{(r+r_+)^2(r^4+a_0^2)}{(r-r_+)(r-r_-)r^4}.\\
\ea Here $r_+=\frac{2M}{(1+P)^2}$, $r_-=\frac{2MP^2}{(1+P)^2}$ and $r_*=\sqrt{r_+r_-}=\frac{2MP}{(1+P)^2}$. Moreover, $a_0=\frac{\Delta}{8\pi}$ with $\Delta$ being the minimal area predicted by LQG \cite{Thiemann:2007pyv} and $P$ is regularization parameter depends on small $\delta\ll 1$ as \ba
P=\frac{\sqrt{1+\gamma^2\delta^2}-1}{\sqrt{1+\gamma^2\delta^2}+1}.
\ea It is clear that when $a_0=P=0$ {(the quantum correction disppears)}, the above
solution \eqref{SelfdualSch} reduces to the Schwarzschild black hole exactly.

\subsection{$\bar{\mu}$-scheme}

Since $\mu_0$-scheme is unphysical and would lead
to wrong semiclassical behavior. To cure this problem, the more complicated quantization scheme of "improved"
dynamics which is usually referred to as $\bar{\mu}$-scheme was formulated. In
this scheme, the quantum parameters $\delta_b$ and $\delta_c$ are chosen as adaptive discreteness variables. The
$\bar{\mu}$-scheme quantization dynamics was first developed in LQC \cite{Ashtekar:2006wn} and later generalized for Kantowski-Sachs models \cite{Chiou08b}.

In this $\bar{\mu}$ quantization scheme, {one} has two choices for {quantum parameters} $\delta_b$ and $\delta_c$. The first choice is the so-called $\bar{\mu}$-scheme with
\ba
\delta_b=\sqrt{\frac{\Delta}{p_b}}    \quad \delta_c=\sqrt{\frac{\Delta}{p_c}}.
\ea
While the second choice is that usually referred {to }as $\bar{\mu}'$-scheme as
\ba
\delta_b=\sqrt{\frac{\Delta}{p_c}}   \quad \delta_c=\sqrt{\frac{\Delta p_c}{p_b}}.
\ea

Though $\bar{\mu}$ and $\bar{\mu}'$-scheme has some advantages over $\mu_0$-scheme, it is still undesire due to it has a large quantum correction at the horizon where usually expected to have little quantum {gravity }influence.

However, as suggested by some authors \cite{Ewing20}, it is maybe not suitable to use
the $\bar{\mu}$-scheme near the horizon because the spatial coordinate becomes null at
the horizon due to the fact that the physical length along that coordinate
in this case will tend to 0. Hence they suggest implementing the $\bar{\mu}$-scheme in terms of another set of coordinates that will not become
null at a horizon. {By} maintaining the areal gauge
for $p_c$ in the classical theory and using the Painleve-Gullstrand coordinates, they obtain a quantum corrected Schwarzschild metric as
\ba
\dd s^2=- f(\tau)\dd \tau^2+ f^{-1}(\tau)\dd x^2+\tau^2\dd\Omega^2,\label{modifiedSch}
\ea where $f(\tau)=1-\frac{2GM}{\tau}+\gamma^2\Delta\frac{4G^2M^2}{\tau^4}$. This modified metric has the correct classical limit at the large distance. The quantum correction decays vary rapidly at large distances, and curvature scalars $R$ are bounded by the Planck scale which is independent of the black hole mass $M$ \cite{Ewing20}. Moreover, this form of the metric was also obtained by other methods such as the quantum Oppenheimer-Snyder collapsing model \cite{Yang22}

Moreover, an attractive polymerized black hole \cite{Effective_bodendorfer_2019,Mass_bodendorfer_2021} solution belongs to a specific $\bar{\mu}$-scheme is constructed recently, in this scheme, the quantum regularization parameters are chosen as
	\begin{eqnarray}
		\delta_b=\pm \frac{4 \lambda_j}{\gamma \left|p_b \right|},\quad \delta_c=\pm \frac{8 \lambda_k}{\gamma \sqrt{\left|p_c \right|} },
	\end{eqnarray} where $\lambda_j$ and $\lambda_k$ are the regularization constants that are related to the inverse Planck curvature and Planck length after rescaling of the fiducial cell \cite{Effective_bodendorfer_2019}.
 This solution leads to quantum extensions
of the Schwarzschild black hole. The quantum corrected Schwarzschild-like metric in this scheme reads \cite{Effective_bodendorfer_2019, Mass_bodendorfer_2021, Testing_brahma_2021,Zhang22c},
\begin{eqnarray}
	\label{mLQG}
	ds^2= -8 A F(r)M^2   dt^2+\frac{dr^2}{8 A M^2 F(r)}+ G(r) (d\theta^2 + \sin^2(\theta) d\phi^2).\label{BBsch}
\end{eqnarray}
where
\begin{eqnarray}
	F(r) &=& \frac{1}{G(r)} \left(\frac{r^2}{8 A  M^2}+1\right) \left(1-\frac{2 M}{\sqrt{8 A M^2+r^2}}\right),\\
	G(r) &=&\frac{512 A^3 M^6+\left(\sqrt{8 A M^2+r^2}+r\right)^6}{8 \sqrt{8 A M^2+r^2} \left(\sqrt{8 A M^2+r^2}+r\right)^3},
\end{eqnarray}
here $M$ represents the mass of asymptotically Schwarzschild black hole, the dimensionless parameter  $A$ is given by $A=\frac12(\lambda_k/M^2)^{2/3}$. {This} type of {quantum corrected} Schwarzschild black hole also admits a black hole to withe hole a bounce \cite{Effective_bodendorfer_2019, Mass_bodendorfer_2021}. In addition, up to now the black hole found in universe are typically with rotations. To this aim, The rotational extension of \eqref{BBsch} through the Newman-Janis algorithm {is} found by authors in \cite{Testing_brahma_2021}.

\subsection{generalised $\mu_0$-scheme}

By considering these weaknesses of $\mu_0$ and $\bar{\mu}$ schemes. Recently, the authors proposed a new scheme \cite{AOS1,AOS2}, which usually referred to as the Ashtekar-Olmedo-Singh (AOS) approach \cite{AOS1,AOS2,AOS3,AOS4,Ashtekar2020}. The AOS scheme can be regarded as an average of the $\mu_o$-type and $\bar\mu$-type schemes. The quantum regularization parameters $\delta_b$ and $\delta_c$ are set to be Dirac observables, i.e. $\delta_b$ and $\delta_c$ being constants along each dynamics trajectory but still may be allowed to vary from one to another. This scheme provides a viable effective description for the Kruskal extension of both the exterior and interior of the Schwarzschild black holes with large mass. This effective dynamics on one hand resolves the interior singularity Schwarzschild black holes in the Planck region while on the other hand keeping the classical horizon regime unchanged and thus overcoming the drawback of the usual $\bar{\mu}$ scheme. By requiring the physical area of the transition surface being the minimal area $\Delta$, they fix the $\delta_b$ as
\begin{equation}\label{eq:deltab}
\delta_b= \left(\frac{\sqrt{\Delta}  } {\sqrt{2\pi} \left(\gamma\right)^{2}GM }\right)^{\frac{1}{ 3}}.
\end{equation}
Meanwhile, $\delta_c$ takes
\begin{equation}\label{eq:deltac}
\begin{aligned}
  \delta_cL_0=\left(\left(\frac{\Delta}{2\pi}\right)^{2} \frac{\gamma}{ 8GM}\right)^{\frac{2}{3}}.
 \end{aligned}
\end{equation} The method of AOS approach has been generalized to Janis-Newmann-Winicour spacetime \cite{Zhang2020}. In these quantum models, the Kretschmann scalar $K=\frac{144M^2}{p_c^3}$ \cite{AOS1,AOS2} is uniformly bounded as
\begin{equation}
\begin{aligned}
K=R_{abcd}R^{abcd} \leq \frac{768 \pi^2}{\gamma ^4 \Delta^2 }.
\end{aligned}
\end{equation}
 The {corresponding} quantum corrected schwarzschild metric in this scheme reads
\ba
\dd s^2=- G(r)\dd t^2+ F(r)\dd r^2+H(r)\dd\Omega^2,\label{AOSSch}
\ea where \ba
G(r)&=&\left(\frac{r}{r_s}\right)^{2\epsilon}\frac{(1-(\frac{r_s}{r})^{1+\epsilon})(2+\epsilon+\epsilon(\frac{r_s}{r})^{1+\epsilon})^2
((2+\epsilon)^2-\epsilon^2(\frac{r_s}{r})^{1+\epsilon})}{16(1+\epsilon)^4(1+\frac{\delta^2_cL^2_o\gamma^2r_s^2}{16r^4})},\\
F(r)&=&\left(1+\frac{\delta^2_cL^2_o\gamma^2r_s^2}{16r^4}\right)\frac{(\epsilon+\left(\frac{r}{r_s}\right)^{1+\epsilon}(2+\epsilon))^2}{(\left(\frac{r}{r_s}\right)^{1+\epsilon}-1)
(\left(\frac{r}{r_s}\right)^{1+\epsilon}(2+\epsilon)^2-\epsilon^2)},\\
H(r)&=&r^2\left(1+\frac{\delta^2_cL^2_o\gamma^2r_s^2}{16r^4}\right).
\ea Here $r_s=2GM$ denotes the Schwarzschild
radius and \ba
\epsilon=\sqrt{1+\gamma^2\left(\frac{\sqrt{\Delta}  } {\sqrt{2\pi} \left(\gamma\right)^{2}GM }\right)^{2/3}}-1.
\ea When quantum correction vanished, i.e $\epsilon=\delta_b=\delta_c=0$, it is clear, that the metric \eqref{AOSSch} goes back to the Schwarzschild solution. It should be mentioned that this model suffers some {criticisms} \cite{Bodendorfer19}, and the reply of these {criticisms} can be found in \cite{Ashtekar2020}.

\subsection{quantum Oppenheimer-Snyder collapsing model}

The gravitational collapse process plays an important role in understanding of the formation black holes. In \cite{Yang22}, the authors propose a quantum Oppenheimer-Snyder gravitational collapse model. First, we note that the interior region $M^-$ consists of a dust ball. This region can be described by Fridmann-Robertson-Walker (FRW) cosmological model. we can write down the flat FRW cosmological line element as \ba
ds^2_{in}=-d\tau^2+a^2(\tau)\left(dr^2+r^2d\Omega^2\right).
\ea Consider a constant $r=r_0$ slice, the induced line element on this slice read \ba
ds^2_{in}=-d\tau^2+a^2(\tau)r_0^2d\Omega^2.
\ea For outside region $M^+$, we assume that this stationary metric can be expressed as \ba
ds^2_{out}=- f(r)\dd t^2+ g^{-1}(r)\dd r^2+r^2\dd\Omega^2.
\ea Again, the induced metric on constant $r$ slice reads
\ba
ds^2_{out}=- (f\dot{t}-g^{-1}\dot{r}^2)\dd \tau^2+r^2(\tau)\dd\Omega^2,
\ea here a dot means derivative with respect to $\tau$. Since the collapse is spherical, the interior
of Schwarzschild {black hole} with the Kantowski-Sachs spacetime is then isometry to each other. By matching the exterior effective spacetime with that the interior effective LQC model describes the interior Kantowski-Sachs spacetime. They found that \cite{Yang22}\ba
g(r)=1-H^2r^2,
\ea where $H=\frac{\dot{a}}{a}$ is the Hubble parameter. By using the classical Fridmann equation $H^2=\frac{8\pi G}{3}\rho$, we have $g(r)=1-\frac{2GM}{r}$ with $M=\frac{4\pi}{3}\rho r^3$ being the total mass of the collapsing ball. While, if we consider the LQC corrected Fridmann equation \ba
H^2=\frac{8\pi G}{3}\rho\left(1-\frac{\rho}{\rho_c}\right).
\ea Then $g(r)=1-\frac{2GM}{r}+\gamma^2\Delta\frac{4G^2M^2}{\tau^4}$ which takes the same form as that in \eqref{modifiedSch}. Moreover, this kind of metric is also obtained \cite{Li2023} by investigating the gravitational collapse of a homogeneous Gaussian dust cloud.

\section{Physical applications}\label{section4}

At the present stage, though there is no completely satisfactory unique solution even for loop quantum corrected Schwarzschild black hole. A lot of physical applications are already carried out based on the current version of loop quantum black hole models. This is done especially in its quantum corrected metric formalism, such as quasi-normal modes \cite{quasinormalmodes1,quasinormalmodes2,Yang22,consistent_bouhmadi-lopez_2020}, black hole shadow \cite{quasinormalmodes1,Yang22}, gravitational lens \cite{Gravitational_fu_2021} and solar system test \cite{Observational_zhu_2020,Zhang22c} or even galaxy scale test for loop quantum black hole. These physical applications can constrain the quantum parameter and therefore help us to confirm or rule out some models. For example, recently a polymer quantum black hole in a $\bar{\mu}$ generalized scheme was proposed \cite{Effective_bodendorfer_2019,Testing_brahma_2021}. However, in this model, the quantum parameters can not be fixed by theoretical considerations. While in \cite{Zhang22c}, the solar system test was applied for such a loop quantum black hole model and the quantum parameter then can be constrained.

Moreover, the Hawking
radiation process was calculated in \cite{Pullin14}. It was found that the discrete quantum geometry introduces a correction much smaller than the leading contribution for large black holes. In addition, using the effective metric \eqref{AOSSch}, the resulting Hawking temperature of this quantum corrected black hole horizon also receives a small quantum correction as \cite{Ashtekar2020}
\ba
T_H=\frac{\hbar}{8\pi K M}\frac{1}{1+\epsilon_m},
\ea where \ba
\epsilon_m=\frac{1}{256}\left(\frac{\gamma\Delta^{1/2}}{\sqrt{2\pi}M}\right)^{\frac{8}{3}}.
\ea

\section{Discussion and outlook}\label{sec:con}
We present a summary of loop quantum Schwarzschild models in {the present} paper, and both the classical and the effective quantum dynamics are considered. We present a brief overview of these topics. We start with the classical Hamiltonian framework of the Schwarzschild black hole spacetime. Then to gain a better understanding of quantum dynamics, we discuss the quantum effective dynamics with four types of quantum regularization parameters $\delta_b$ and $\delta_c$. These four types of parameters are commonly used for describing quantum dynamics. All the corresponding quantum corrected models can resolve the singularity inside the Schwarzschild black hole. However, they also exhibit some significant differences. Moreover, we also summarize the physical applications of these models.

The field of loop quantum black holes now is still in its infancy stage. Unlike LQC where one has a unique way to fix the quantum parameters. At the present stage, there is still no completely satisfactory unique solution even for a Schwarzschild black hole. Moreover, there are still many open issues in the current paradigm, for example, how to realize the Hawking radiation in such a loop quantum black hole scenario?
Can we resolve the black hole information paradox in the loop quantum gravity setup? Moreover, in loop quantum black hole models, the general covariance of spacetime is often discussed. Some authors stated that covariance cannot be addressed in spherical symmetrical models because of slicing dependence \cite{Bojowald15}. The issue is currently still an open issue and under debate. Some possible general covariance model is constructed \cite{GC1,GC3} and the comments on these models can be found in \cite{GC2}. These open questions require a deeper understanding of the loop quantum black hole model.

\begin{acknowledgements}
This work is supported by NSFC with No.12275087, No. 11775082, and ``the Fundamental Research Funds for the Central Universities''.

\end{acknowledgements}

\end{document}